\documentclass[aps,prb,twocolumn,groupedaddress,showpacs,reprint]{revtex4-1}
\usepackage[english]{babel}
\usepackage[utf8x]{inputenc}
\usepackage{amsmath,amssymb,amstext}
\usepackage{mathrsfs}
\usepackage{graphicx}
\usepackage{color}

\newcommand{\ket}[1]{\left\vert #1\right\rangle}

\newcommand{\eps}{\varepsilon}

\newcommand{\be}{\begin{equation}}
\newcommand{\ee}{\end{equation}}
\newcommand{\bea}{\begin{eqnarray}}
\newcommand{\eea}{\end{eqnarray}}
\newcommand{\tr}[1]{\mathrm{Tr} \left [ #1 \right ]}

\begin{document}

\title{Theoretical investigation of spin-filtering in CrAs / GaAs heterostructures}
\author{B. A. Stickler}
\email{benjamin.stickler@uni-graz.at}
\affiliation{Institute of Physics,
 Karl-Franzens Universit\"{a}t Graz, Austria}
\author{C. Ertler}
\affiliation{Institute of Physics,
 Karl-Franzens Universit\"{a}t Graz, Austria}
\author{L. Chioncel}
\affiliation{Theoretical Physics III, Center for Electronic Correlations and Magnetism, Institute of Physics, University of Augsburg,
D-86135 Augsburg, Germany}
\author{W. P\"{o}tz}
\affiliation{Institute of Physics,
 Karl-Franzens Universit\"{a}t Graz, Austria}

\begin{abstract}
The electronic structure of bulk fcc GaAs, fcc and tetragonal CrAs, and CrAs/GaAs supercells, computed within LMTO local spin-density functional theory, is used  to extract the band alignment (band offset) for the  [1,0,0] GaAs/CrAs interface in dependence of the spin orientation.  With the lateral lattice constant fixed to the experimental bulk GaAs value, a local energy minimum is found  for a tetragonal CrAs unit cell with a slightly ($\approx$ 2 \%) reduced longitudinal  ([1,0,0]) lattice constant.  Due to the identified spin-dependent band alignment, half-metallicity of CrAs no longer is a key requirement for spin-filtering.  Encouraged by these findings, we study the spin-dependent tunneling current in  [1,0,0] GaAs/CrAs/GaAs heterostructures within the non-equilibrium Green's function approach for an effective tight-binding Hamiltonian derived from the LMTO electronic structure.  Results indicate  that these heterostructures are probable candidates for efficient room-temperature all-semiconductor spin-filtering devices.   
\end{abstract}

\maketitle

\section{Introduction}

The identification and design of high-efficiency all-semiconductor spin-filtering devices which operate at room temperature and zero external magnetic field are of profound interest for spintronic applications.\cite{zutic04,fabian07}  In spite of some success, for example regarding improved spin injection from ferromagnets into Si, progress has been moderate. Diluted magnetic semiconductors, such as GaMnAs, which have been grown most successfully with good interfacial quality onto nonmagnetic fcc semiconductors, generally, have critical temperatures far below room temperature and are hampered by disorder.\cite{burch08,sato10,ertler12}   Growing high-quality GaMnAs layers within heterostructures proves to be more difficult than in the bulk.
% Most likely, due to structural problems in the form of unwanted defects leading to free-carrier compensation, an inevitable disorder in ternary alloys, and limitations to the hole doping levels in the semiconductor contact layers,  in experiment,  ferromagnetic order appears to be difficult  to establish and to maintain under bias in AlGaAs/GaMnAs heterostructures.\cite{ohya11,ertler12}
Signatures of weak quantum confinement effects associated with GaMnAs quantum wells in GaAlAs/GaMnAs single- and double barrier heterostructures have been reported in the literature. \cite{ohya11} 
Using one Ga$_{0.96}$Mn$_{0.04}$As contact layer on an asymmetric GaAlAs / GaMnAs heterostructure magnetization-dependent negative-differential-resistance has been observed.\cite{likovich09}  A similar TMR experiment has been performed recently, in which resonant tunneling in non-magnetic 
AlGaAs/GaAs/AlGaAs was used to control the hole current.\cite{muneta12}  
 
It is an experimental fact that bulk MAs or  bulk MSb compounds, with M denoting a transition metal, such as V, Cr, and Mn,  do not have their ground state in the fcc phase.  Even as thin layers on an fcc substrate, such as GaAs, they appear to be difficult to grow in a lattice-matched form.\cite{sanvito00,zhao05,hashemifar10,mavropoulos07}  Important exceptions are the reports of the experimental realization of fcc MnAs quantum dots on GaAs and the hetero-epitaxial growth of thin layers of  CrAs  on GaAs substrates.\cite{ono02,akinaga00,mizuguchi02,mizuguchi02b,mizuguchi02b,mavropoulos07}
Recent experiments have suggested that CrAs can be grown epitaxially in the fcc structure on top of GaAs and displays ferromagnetic behavior well above room temperature.\cite{akinaga00,mizuguchi02, bi06} Ab-initio studies of the system have revealed that fcc CrAs is a half-metallic ferromagnet and have led to the prediction that the Curie temperature of fcc CrAs may be as high as $1000$ K.\cite{kuebler03,galanakis11,mavropoulos07,galanakis03} 
The electronic structure of GaAs / CrAs heterostructures and  transport properties through GaAs/CrAs/GaAs tri-layers have been studied by Bengone {\it et al.}\cite{bengone04}.

Recently  the effect of electronic correlations upon the half-metallicity of stacked 
short period (CrAs)$_\ell$/(GaAs)$_\ell$ ($\ell \le 3$) superlattices along [001] has been investigated.   Results indicate that 
the minority spin half-metallic gap is suppressed by local correlations at finite temperatures and continuously 
shrinks on increasing the heterostructure period. As a consequence, at the Fermi level, the polarization 
is significantly reduced, while dynamic correlations produce only a small deviation in magnetization~\cite{chioncel11}. 
Essentially the same effect was found for defect-free digital magnetic heterostructures $\delta$-doped 
with Cr and Mn~\cite{chioncel11b}. In addition both studies indicate that  the minority 
spin highest valence states below the Fermi level are localized more in the GaAs layers while the lowest conduction band states 
have a many-body origin derived from CrAs. Therefore independent of the presence of electronic correlations in these 
heterostructures holes and electrons may remain separated among different layers which may be detected in photo-absorption 
measurements.

Another density-functional theory (DFT) based calculation, as well as, an experimental 
report have lead to the claim that
the fcc-structure of thin film CrAs is energetically unstable.\cite{etgens04,hashemifar10}  It was argued that the experimentally observed  
ferromagnetic behavior reported in Refs. \onlinecite{akinaga00,mizuguchi02, bi06} may be caused by magnetic defects near the heterointerface and that no 
half-metallicity may be present at all. An additional problem to our knowledge,  not studied in detail as of yet,  may be the uncontrollable diffusion of Cr into GaAs thereby forming deep traps.\cite{privcommstrass}. It has to be noted that the final answer to the structural properties and stability of thin films of CrAs on top of GaAs, as well as CrAs/GaAs heterostructures,  can only be given by or in conjunction with further experiments. 
It is indeed the aim of this article to stimulate these experimental studies by demonstrating that epitaxial CrAs/GaAs heterostructures should function as very efficient all-semiconductor room-temperature spin-filtering devices.

Exploring spin-filtering in GaAs - CrAs - GaAs  heterostructures,  first we investigate the stability of the CrAs unit cell under longitudinal  distortion. 
It is a central result of this work that a tetragonal, local-minimum bulk phase of CrAs can be found which is very near to being fcc lattice matched to bulk GaAs.  This implies that one may  perform a transport calculations under the assumption of perfect fcc lattice matching without introducing a large systematic error.  The band alignment is determined with the help of two different approaches, which both yield rather similar results. This provides a reasonable error estimate for the method employed, and leads us to the conclusion that half-metallicity is not a necessary ingredient for efficient spin-filtering. Rather, the interfacial properties and the spin-selective band alignment between the CrAs and GaAs layers appear to be essential.  We map the electronic structure of the bulk materials onto an effective $20$-orbital sp$^3$d$^5$s$^*$ nearest-neighbor empirical tight-binding (ETB) model, which is particularly suited for non-equilibrium transport 
calculations.\cite{dicarlo2003,stovneng94,datta,boykin91} In a final step we calculate the spin-selective current-voltage (I-V) characteristics within the non-equilibrium Green's function formalism for different layer thicknesses and temperatures.

This article is structured as follows.  In Sec. \ref{sec:lmto} we summarize the theoretical approach and present  the results for the electronic structure within the LMTO method.  In Sec. \ref{sec:tb} we describe  the mapping procedure onto the tight-binding model.  The transport model and results for the spin-current response are discussed in Sec. \ref{sec:trans}.  Summary and conclusions are given in Sec. \ref{sec:conc}.

\section{LMTO electronic structure calculations and the fcc CrAs / fcc GaAs  [1,0,0]  band offset} \label{sec:lmto}

The electronic structures of bulk zinc-blend (ZB) GaAs, bulk ZB and tetragonal CrAs, as well as that of several CrAs / GaAs supercells, have been determined employing the LMTO-ASA code, as developed by Jepsen and Andersen.\cite{lmtocode,skriver}  A closely related code has been used previously  to explore the electronic structure of bulk ZB CrAs and thin-layer fcc GaAs / CrAs superlattices.  In particular the half-metallic behavior as a function of superlattice period and lattice constants has been investigated.\cite{chioncel05,chioncel11} Details of this approach, as well as its benefits and caveats, can be found in the literature.\cite{skriver}

The electronic structure model is based on the local spin-density approximation (LSDA) omitting spin-orbit interaction and corrections for strong correlations  as provided, for example, by the dynamic-mean-field-theory\cite{georges96} or the variational cluster approach\cite{gros93,potthoff03}. However, it has been shown by Chioncel et al.\cite{chioncel05,chioncel11} that the inclusion of correlations does not affect the magnetization and only leads to minor corrections to the band structure. It is well known that LSDA or its gradient-corrected approximation produce band gaps that are typically at least $30\%$ smaller than the experimental values for almost all semiconductors and insulators.  In our case, too,  the experimentally verified GaAs band gap of 1.52 eV at low temperatures is strongly underestimated by the present DFT method, predicting a value of about  0.35 eV. However, the overall features in the vicinity of the band gap are reproduced reasonably well.  Since we do not study transport across the 
main energy 
gap, such 
as in Zener tunneling,  this shortcoming is without any further disadvantage. Therefore, in what follows we consider two distinct cases: (i) we leave the electronic structure unchanged, i.e. we use the GaAs band structure with the underestimated band gap, and (ii) we scissor the GaAs electronic structure  to the 
experimentally observed value of $1.52$ eV. As we shall see, this does not affect the statement that CrAs / GaAs heterostructures function as spin-filters. Furthermore, we consider n-doped GaAs such that the spin-orbit effects, at least in GaAs, can be neglected. In the case of half-metals one may speculate that the problems of LSDA are not as serious, since the dielectric response of half-metals is of metallic type. Previous results have shown that the detailed nature of states around $E_F$ in bulk CrAs is changed when different lattice constants or models for electronic correlations are considered, while the overall features of the electronic structure are preserved.\cite{chioncel11b}   Clearly the final answer as to the accuracy of the electronic structure model can only be given by comparison to experiment. 

Since the GaAs band structure is well known and, except for the band gap, is fairly well reproduced within the  LMTO-ASA code, here we confine ourselves to results obtained for CrAs. In a first approach we assume that the fcc CrAs lattice constant is equal to the GaAs lattice constant $a_{\rm GaAs}=5.65$ \AA.  This choice is motivated by the hope that thin layers of CrAs can be grown onto a GaAs substrate in lattice-matched fashion.  The electronic structure for majority and minority carriers  is shown, respectively,  in Figs. \ref{fig:esmaj} and \ref{fig:esmin} (red solid lines).  For the minority carriers a gap of about $1.8$ eV is predicted. A more detailed analysis of the electronic structure shows the origin of the valence band edge in the As-p - Cr-d hybrid orbitals, while conduction band edge states  are dominated by the Cr-d orbitals, see Chioncel et al. \cite{chioncel05,chioncel11}. Further, we observe from the ab-initio calculations that the lattice constant chosen for bulk fcc CrAs determines 
whether CrAs is half-metallic or not, see also \cite{chioncel05,chioncel11}. Since the bulk band structures of GaAs and CrAs combined in a heterostructure are 
shifted relative to each other by the band offset, as discussed below, we are led to the conclusion  that half-metallicity of the CrAs layer is not compulsory for the realization of a spin-filter.  A necessary ingredient is a highly
spin-dependent  CrAs  band structure.

Moreover, we investigate the total energy of the bulk CrAs unit cell as a function of the longitudinal, i.e. the out-of-the-plane, lattice constant $a_\perp$ while leaving the in-plane lattice constant $a_{||}$ fixed thus making the unit cell tetragonal. It is reasonable to assume that, for thin lattice-matched layers of CrAs on top of GaAs, the in-plane lattice constant $a_{||} = a_{\rm GaAs}$ takes on the value of bulk GaAs, i.e. that the in-plane symmetry is preserved. It is found, that the minimal energy is achieved for $98$ \% of the GaAs lattice constant, as depicted in Fig. \ref{fig:comp}, i.e. $a_\perp = 0.98 a_{||}$. Hence, the actual unit cell of CrAs on the GaAs [1,0,0] surface is not perfectly cubic but tetragonal. The mere existence of such a (local) minimum close to the GaAs lattice constant is non-trivial and an important result and motivation for the spin transport analysis to follow.   Moreover, a comparison of the bands of fcc CrAs with the bands of tetragonal CrAs with $a_\perp = 0.98 a_{
||}$ reveals that the electronic structure is almost unaffected by this distortion, see Figs. \ref{fig:distmaj} and {\ref{fig:distmin}. Therefore the systematic error introduced by assuming that the CrAs unit cell is perfectly cubic with $a_{||} = a_\perp \equiv a_{\rm CrAs} = a_{\rm GaAs}$ is 
negligible when compared to other uncertainties.  We shall from now on assume that bulk CrAs (on [1,0,0] GaAs) has the fcc crystal structure.

\begin{figure}[h!]
 \centering
 \includegraphics[width = 80mm]{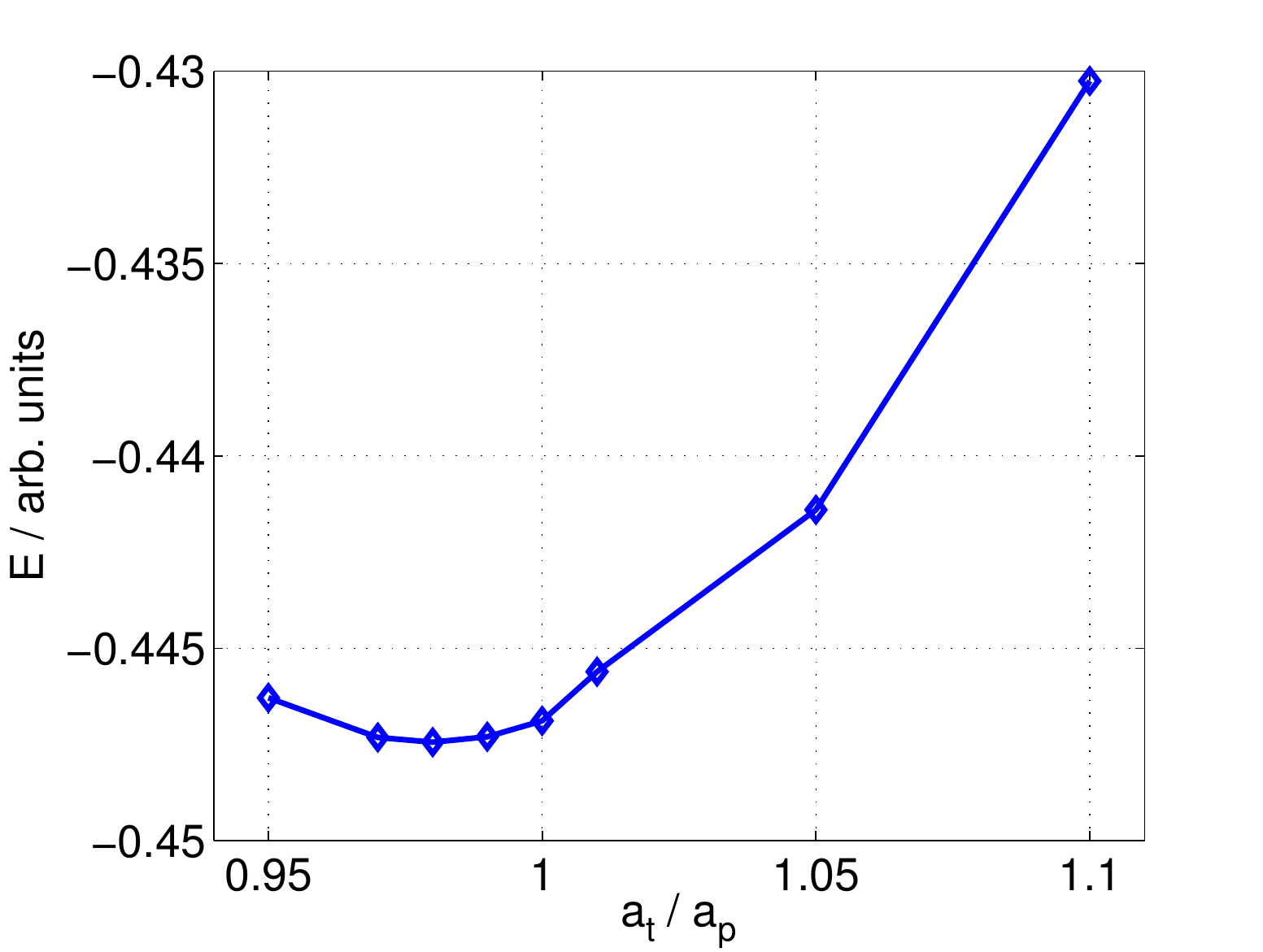}
 \caption{(Color online) Total energy of the bulk CrAs unit cell as a function of lateral lattice constant $a_{\perp} / a_{||}$.} \label{fig:comp}
\end{figure}

\begin{figure}[h!]
 \centering
\includegraphics[width = 80mm]{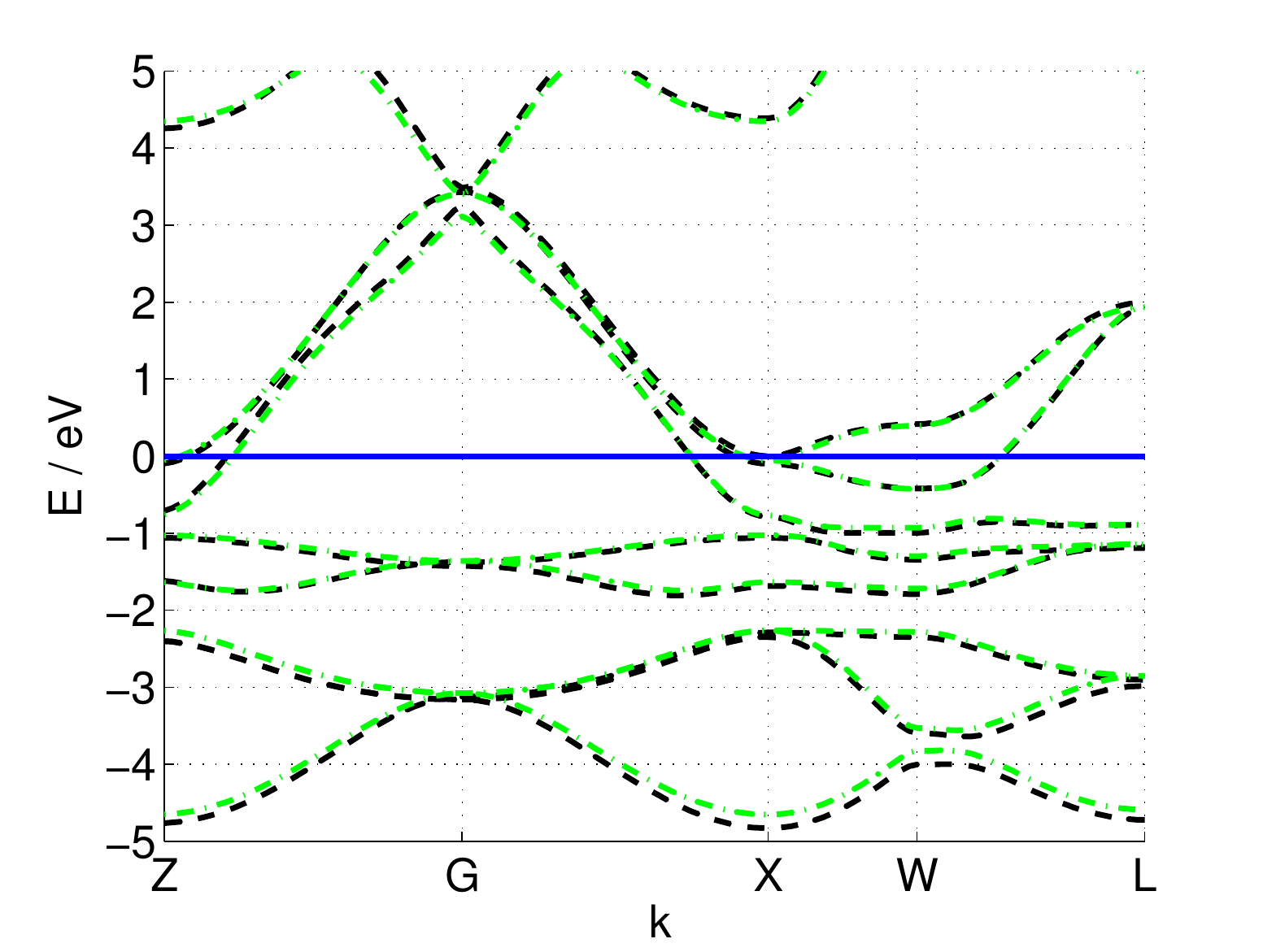}
 \caption{(Color online) Ab-initio electronic structure of majority spin CrAs for the fcc (green solid line) and the tetragonal unit cell (black dashed line). The {\sc Fermi} energy is indicated by a solid blue line.} \label{fig:distmaj}
\end{figure}

\begin{figure}[h!]
 \centering
\includegraphics[width = 80mm]{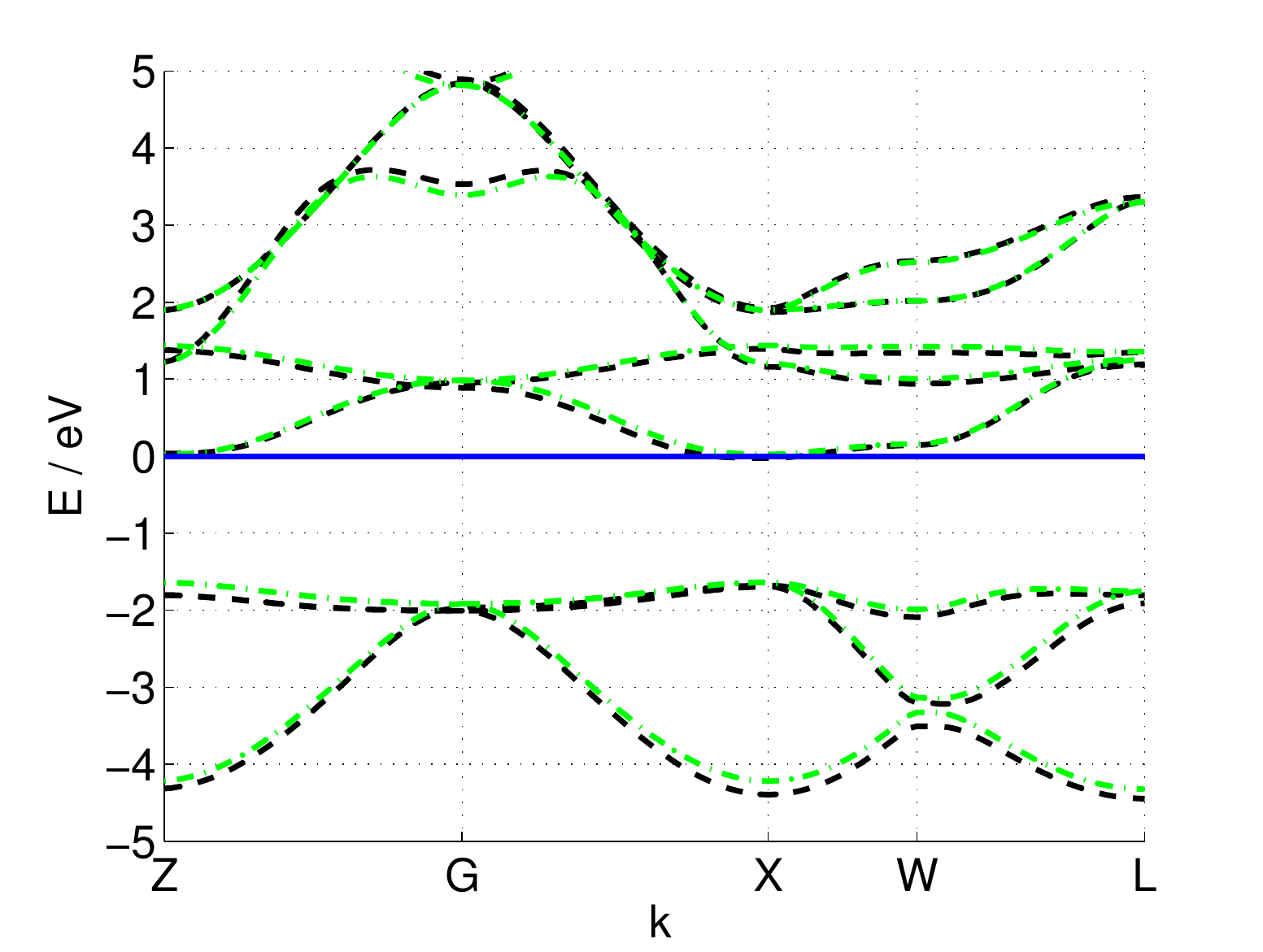}
 \caption{(Color online) Ab-initio electronic structure of minority spin CrAs for the fcc (green solid line) and the tetragonal unit cell (black dashed line). The {\sc Fermi} energy is indicated by a solid blue line.} \label{fig:distmin}
\end{figure}

In the spirit of core level spectroscopy, the band offset at the [1,0,0] GaAs / CrAs interface was determined using the energy of a low-lying reference state that is present in both bulk materials, as well as in the CrAs / GaAs supercell. We chose two different reference states and compared the values obtained with each of them in order to have a reasonable error estimate for the method. The band offset $\Delta$ is calculated according to \cite{wei98}
\be
\Delta = \Delta_G - \Delta_C + \Delta_{GG} - \Delta_{CC}.
\ee

In the first evaluation, we use the low-lying As-4s bands for our reference states. We compute the center of energy of these bands for bulk fcc CrAs $\Delta_C$, bulk fcc GaAs $\Delta_G$, as well as for the As atoms embedded in the Ga and Cr environment in a (GaAs)$_6$(CrAs)$_6$ supercell $\Delta_{GG}$ and $\Delta_{CC}$. All energies are taken relative to the respective Fermi energy. The center of energy of one particular band is obtained by calculating the fat-band weights with the help of the ab-initio LMTO-ASA code and using them for a weight factor when performing the average over the first Brillouin zone.  The second way of calculating the band offset is based on including the As-3d orbitals into the self-consistent loop of the ab-initio calculation. The reference state is then taken as the {\it center of band} as obtained from the LMTO-ASA code.\cite{lmtocode}  In both cases we obtain a band offset $\Delta$ in the range of $\Delta_1= 0.6 \pm 0.2 $ eV, for majority spin, and a band offset of $\Delta_2=0.
5 \pm 0.2$  eV, for minority spin bands.   Although 
this error may seem considerable, in our calculations,  it is of little significance as to the performance of  CrAs / GaAs heterostructures as  a spin-filter.  In the calculations below we chose  $\Delta_1 = 0.60$ eV and $\Delta_2 = 0.55$ eV, respectively, for majority and minority spin, as obtained within the method based on the As-4s reference states.

\begin{figure}[h!]
 \centering
 \includegraphics[width = 80mm]{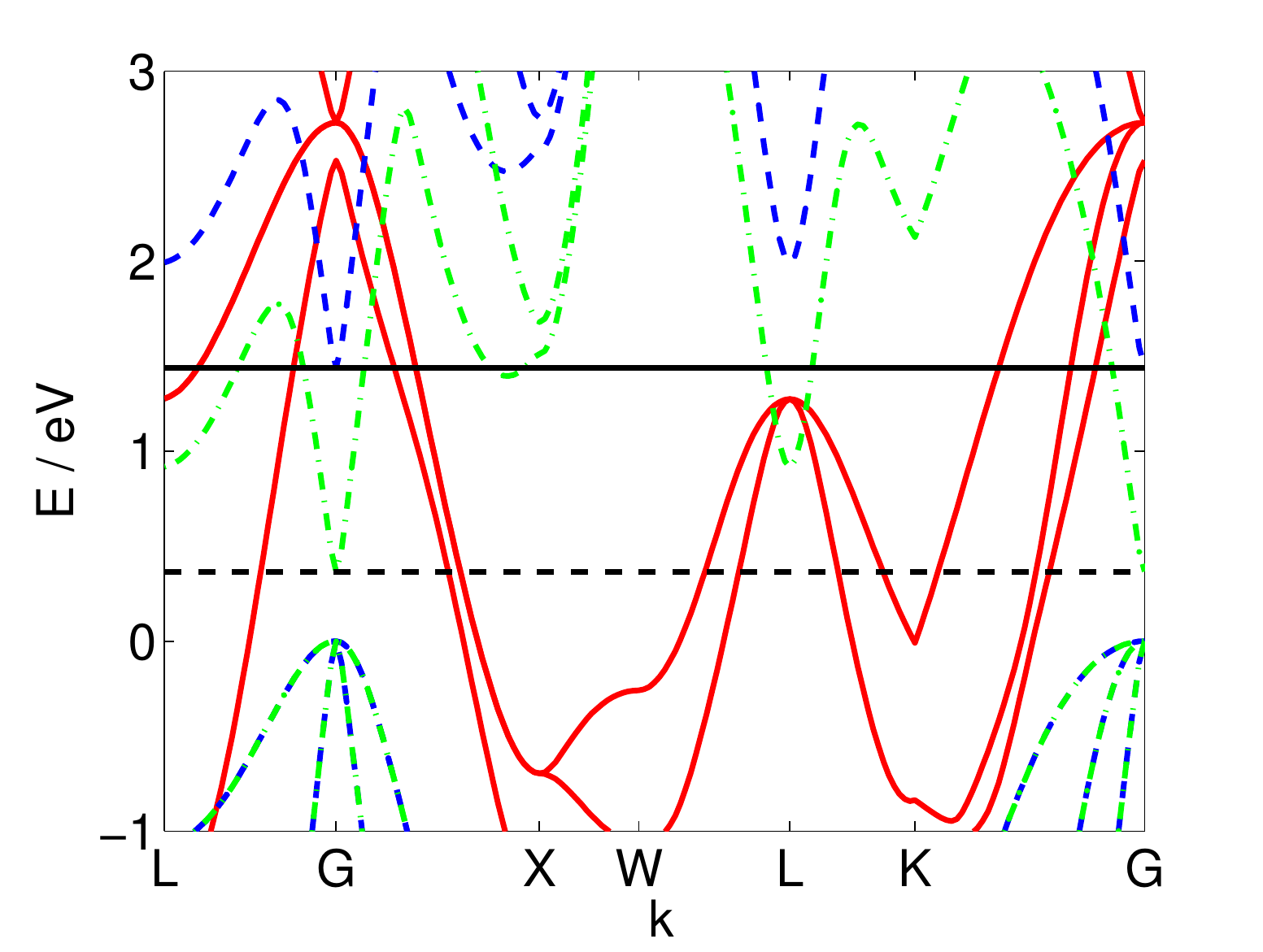}
 \caption{(Color online) Majority spin CrAs ab-initio band structure (red solid line),  scissored, and un-scissored GaAs band structure (blue-dashed, respectively, green-dash-dotted line). The Fermi energy $E_F = 0.01$ eV above the conduction band minimum of GaAs is indicated by the horizontal lines (solid: scissored GaAs, dashed: un-scissored GaAs).} \label{fig:esmaj}
\end{figure}

\begin{figure}[h!]
 \centering
 \includegraphics[width = 80mm]{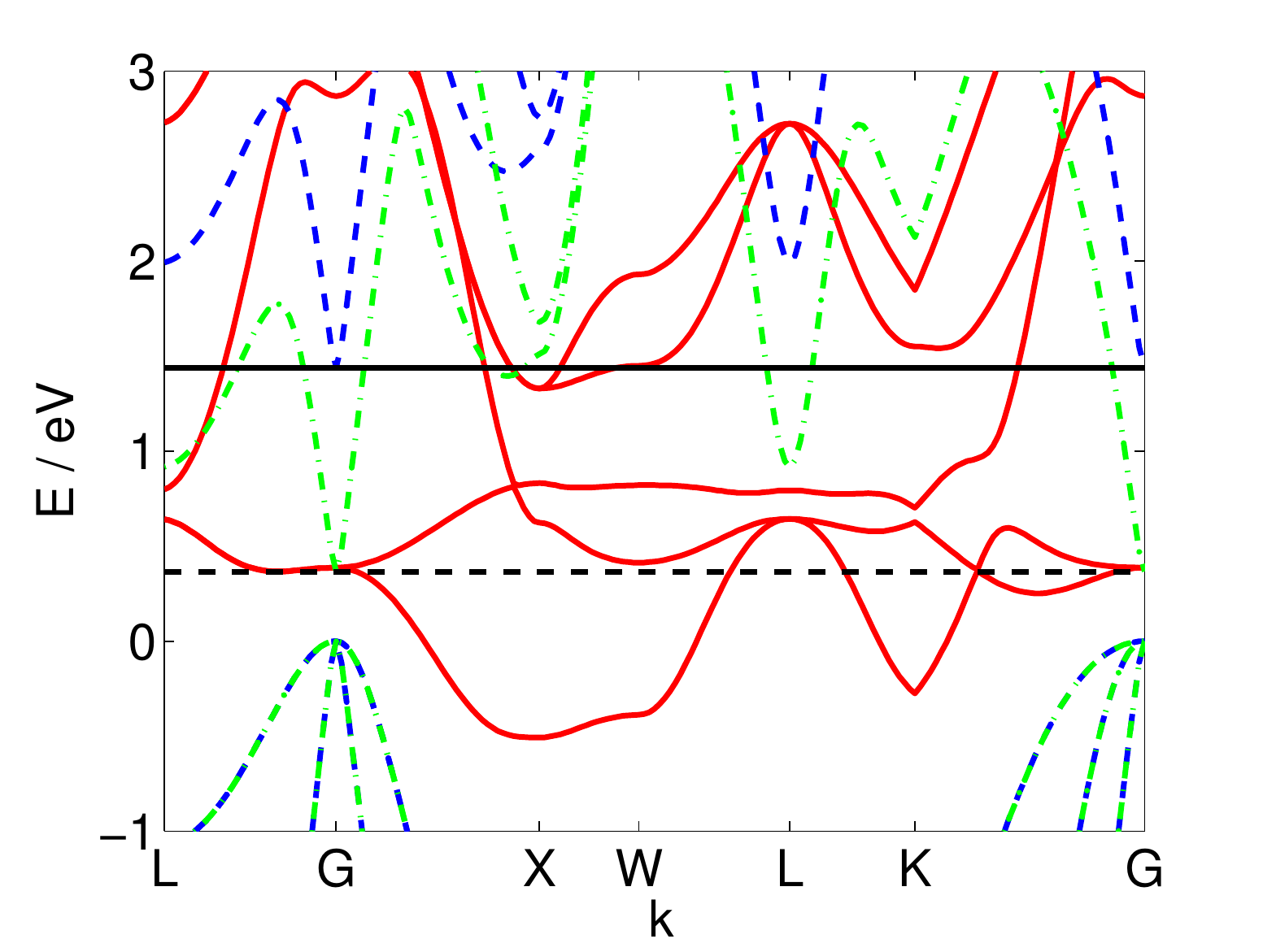}
 \caption{(Color online) Minority spin CrAs ab-initio band structure (red solid line) and scissored as well as un-scissored GaAs band structure (blue-dashed, respectively, green-dash-dotted line). The Fermi energy $E_F = 0.01$ eV above the conduction band minimum of GaAs is indicated by the horizontal lines (solid: scissored GaAs, dashed: un-scissored GaAs).} \label{fig:esmin}
\end{figure}

In Figs. \ref{fig:esmaj} and \ref{fig:esmin} we show, on top of the electronic structure of majority- and minority-spin CrAs,   the electronic structure of GaAs for case (i) (un-scissored gap)  and case (ii) (scissored gap) taking into account the band offset. It can now clearly be seen that carriers injected from n-GaAs near the $\Gamma$-point do not reach the main energy gap region of bulk CrAs.  Even in case of p-doping of the GaAs contact layers, this region can only be reached under an applied bias exceeding approximately $0.3$ V, as can be observed from an inspection of Fig. \ref{fig:esmin}. Moreover, scissoring of the GaAs band gap does not significantly change the spectral overlap of the lowest GaAs conduction band  with the CrAs bands near the $\Gamma$ point.  The situation is rather different for minority carriers, however.  Here the overlap of the lowest GaAs conduction band occurs, in both cases, with fairly flat CrAs bands of low mobility. Hence, the actual value of the band offset is not too 
important which allows the claim that the presence of half-metallicity in the CrAs layer is not mandatory for spin-filtering in CrAs / GaAs heterostructures.   In view of these results and the broad-band form of the electronic structure of CrAs in the relevant energy window, in particular, it appears that spin-filtering is  rather stable with respect to moderate modifications to the value of the band offsets arising, for example,  from interfacial defects or local lattice distortions.

\section{Effective sp$^3$d$^5$s$^*$ tight-binding model} \label{sec:tb}

The results obtained within the LMTO electronic structure investigation encourage one to go one step further and to investigate spin-selectivity in the electric current for CrAs/GaAs heterostructures under bias.  Since the LMTO calculations are based on thermal equilibrium we first map the relevant segments of the electronic structure onto an empirical tight-binding (ETB) model on which we then base  the non-equilibrium transport study.
Specifically, each of the ab-initio electronic structure $\eps_n^{LSDA}$ of GaAs and of majority- and minority-spin CrAs  are mapped onto an effective sp$^3$d$^5$s$^*$ nearest-neighbor ETB model.\cite{slater54,jancu98,starrost95} We chose an ETB model because it is particularly well-suited for non-equilibrium steady state transport calculations with the help of Green's functions.\cite{dicarlo2003,stovneng94,boykin91,datta}

In principle, this step can be avoided if the LSDA wave functions were used to express the transmission function of the heterostructure, as for example proposed within the SIESTA DFT approach.\cite{brandbyge02} The trade-off of an approach which is based on the L(S)DA one-particle wave functions, however, is that its validity ad-hoc is questionable, since the  wave functions used in the Kohn-Sham variational principle do not allow a direct physical interpretation.  Their connection to the S-matrix of the (many-body) system is not obvious. The use of ground state wave-functions definitely limits one to the linear-response regime, since the transmission function would be obtained for zero external bias only. In general, L(S)DA bulk band structure calculations do not produce the correct energy gap.  A simple scissoring strategy cannot be performed for a heterostructure, thus, any deficiencies in the ab-intio electronic structure are carried over inevitably into the transport calculation.  These difficulties 
have convinced us to follow a mapping approach from the ab-initio band structure calculation to an empirical tight-binding model and to take care of the interface problem in an additional step.

A mapping of the electronic structure onto an ETB model does not, in the first place, introduce any further systematic error, as long as the electronic structure in the relevant energy region is well approximated. Moreover, this approach comes with further benefits, such as computational effectiveness.  It does not only allow one to account for known deficiencies of the electronic structure within DFT, such as a  scissoring of the GaAs main band gap to the experimentally verified value, but also to control the particular modeling of the interface (e.g. by the inclusion of defects)  and to treat a genuine non-equilibrium situation self-consistently.  Bias-dependent mean-field corrections or disorder effects can be added in self-consistent fashion, as utilized by some of us recently for similar systems.\cite{ertler10,ertler11} 
 
The formal mapping process for a given bulk material is executed using a genetic algorithm, as implemented in Matlab,  to minimize the cost functional
\be \label{eq:costf}
K(\xi) = \sqrt{\sum_{n{ k}} a_n({ k}) \left [ \eps_n^{LSDA}({ k}) - \eps_n^{ETB}({ k}, \xi) \right ]^2}~. 
\ee
Here, $a_n(k)$ are  normalized weights where $n$ is the band index and $k$ the wave vector, $\xi$ denotes the set of 31 independent ETB parameters in the sp$^3$d$^5$s$^*$ basis\cite{slater54} and $\eps_n^{ETB}({ k}, \xi)$ is the ETB band structure as a function of $k$ and $\xi$. The weights $a_n({ k})$ are used to restrict and focus the fit to the part of the band structure which contributes to charge transport.  This ensures that the energy bands  are well represented by the ETB fit and no  "spurious bands" appear inside a chosen energy window.

Subsequently, the Hamiltonian of a given heterostructure can be put together layer by layer in a straight-forward fashion. This convenient layer-by-layer construction  can be carried over to the construction of the non-equilibrium Green's function components and has been used in the calculations below. A further advantage  is that we know the $k$-dependence of the bulk ETB Hamiltonian matrices analytically via the structure factors.\cite{slater54} Hence, the influence of small deviations in the ETB binding parameters can be investigated systematically.

One problem, however,  arises when the interface between GaAs and CrAs has to be modeled in this modular approach.\cite{toymodel}\footnote{Note that the SIESTA DFT approach, too, employs an ad-hoc approximation at the interface which may be regarded as rather questionable.\cite{brandbyge02}} Let us briefly discuss the problem of how to model the interface in an adequate way, consistent with the available information, which are the ETB bulk Hamiltonians and the band offsets.  The standard approach is to invoke the virtual crystal approximation (VCA).\cite{stovneng94,diventra,dicarlo2003,harrison,lake97,strahberger2000,boykin91} It has been demonstrated that this approximation is, in general,  inconsistent since it violates symmetries underlying  the bulk ETB Hamiltonians.\cite{toymodel}  In particular, the ETB parameters are not uniquely defined by the electronic structure alone, however, the VCA requires a unique identification of the ETB parameters.\cite{toymodel} There are two possible remedies to this 
apparent 
inconsistency. First, we can formulate {\it matching conditions} which are a discrete form of the matching conditions in continuous space quantum mechanics. This approach works for pairs of band-to-band transitions and has been discussed extensively by Stickler and P\"{o}tz \cite{toymodel}. 
Here, in this more complex situation, we shall follow an alternative approach made possible by the presence of As as a common anion: we can ensure that the ETB parameters can be uniquely attributed to one atomic species by posing a further constraint onto the fitting 
procedure Eq. \eqref{eq:costf}: We require the As onsite energies (under zero bias and band offset ) to be constant throughout the device. This is in accordance with ETB theory.\cite{slater54} Hence, the mapping onto a ETB model is executed in two steps: In a first step we fit the majority and minority spin band structure of CrAs independently without any further restrictions. In a second step we fit the GaAs band structure under the constraint that all As onsite energies have to have the same value as in CrAs. Let us briefly discuss the implications of such a fitting procedure: First of all,  we obtain two different sets of ETB parameters for GaAs  since we fit the two CrAs band structures independently and then restrict the GaAs parameters depending on spin orientation. Had we included the spin-orbit interaction into our model, two distinct sets of ETB parameters for GaAs would come more natural, however, in any case the values obtained must be considered best fits under given constraints. Moreover, we 
emphasize that the particular form of the ETB parameters has no influence on the transport as long as the electronic structure including the band offset is reproduced reasonably well.\cite{toymodel}

All in all we perform the down-folding process for four different combinations: (A) majority-spin CrAs  and scissored GaAs, (B) majority-spin CrAs and un-scissored GaAs, (C) minority-spin CrAs and scissored GaAs, and (D) minority-spin CrAs and un-scissored GaAs.    
The ETB parameters which were identified as optimal for each of  the four cases are listed in the Appendix. The fits for GaAs together with the ab-initio band structure and the fits of majority- and minority-spin CrAs  are given in Figs. \ref{fig:fita}, \ref{fig:fitb}, \ref{fig:fitc} and \ref{fig:fitd}.  In view of the fact that the computed ab-initio electronic structure will at best capture the overall features of the actual electronic band structure the ETB fits achieved  are highly satisfactory.

In principle, the spin-orbit interaction can be included in the ETB model following the work of Chadi.\cite{chadi77}  However, since the LMTO-ASA code itself currently does not feature spin-orbit interactions, its implementation at the ETB level would require the introduction of  further (and somewhat arbitrary) parameters into our model, in particular for CrAs.    Furthermore,  we shall focus on transport with n-doped GaAs buffer  layers 
so that a detailed account of the spin-orbit interaction in the GaAs electronic structure will not really be important here.

\begin{figure}[h!]
 \includegraphics[width = 80mm]{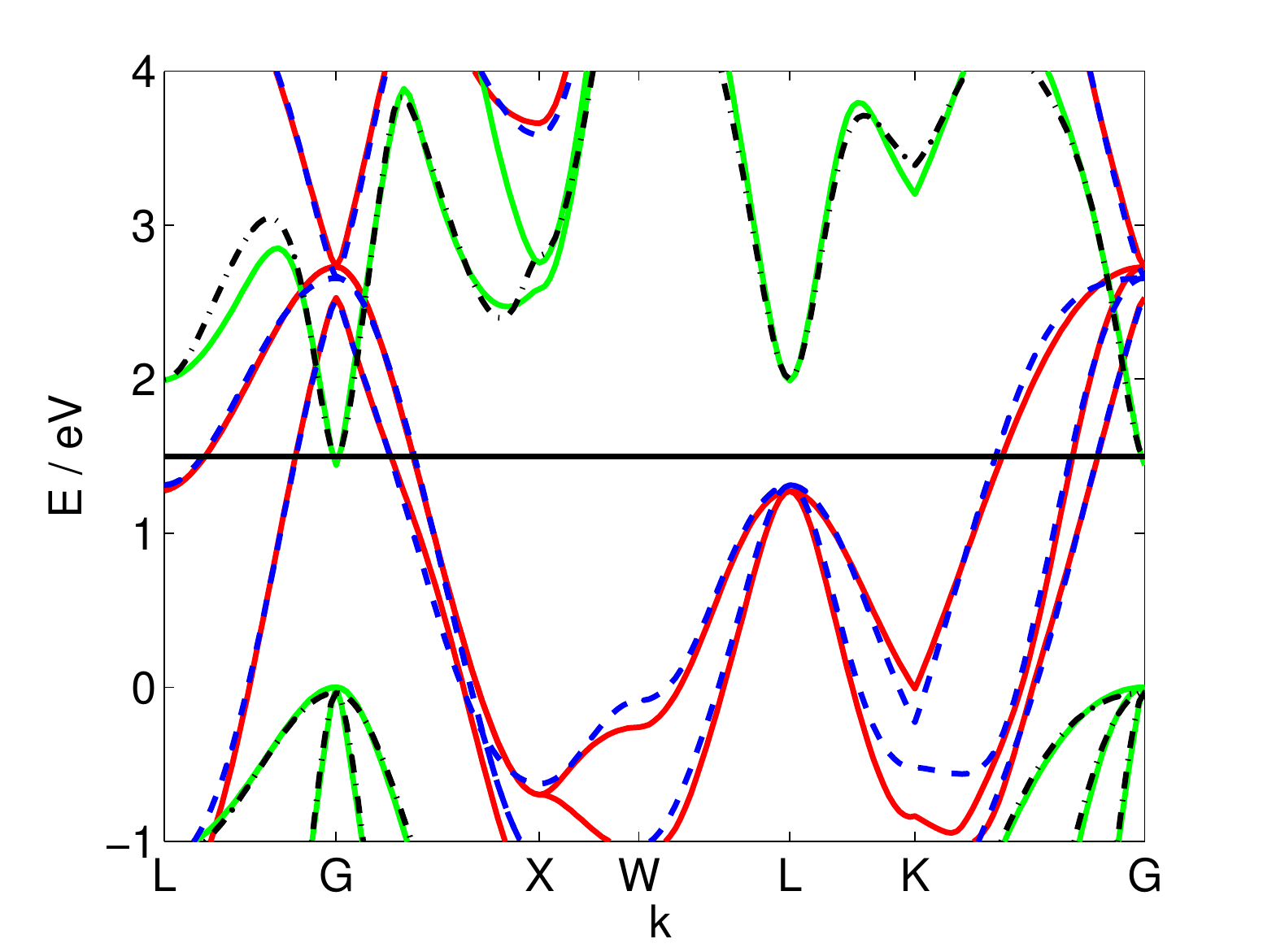}
 \caption{(Color online) Scenario (A): majority spin CrAs ab-initio band structure (red solid line), CrAs ETB-fit (blue dashed line), scissored GaAs band structure (green solid line) and GaAs ETB fit (black dashed line).
The Fermi energy $E_F = 0.01$ eV above the conduction band minimum of GaAs is indicated
by the horizontal solid line.} \label{fig:fita}
\end{figure}

\begin{figure}[h!]
 \includegraphics[width = 80mm]{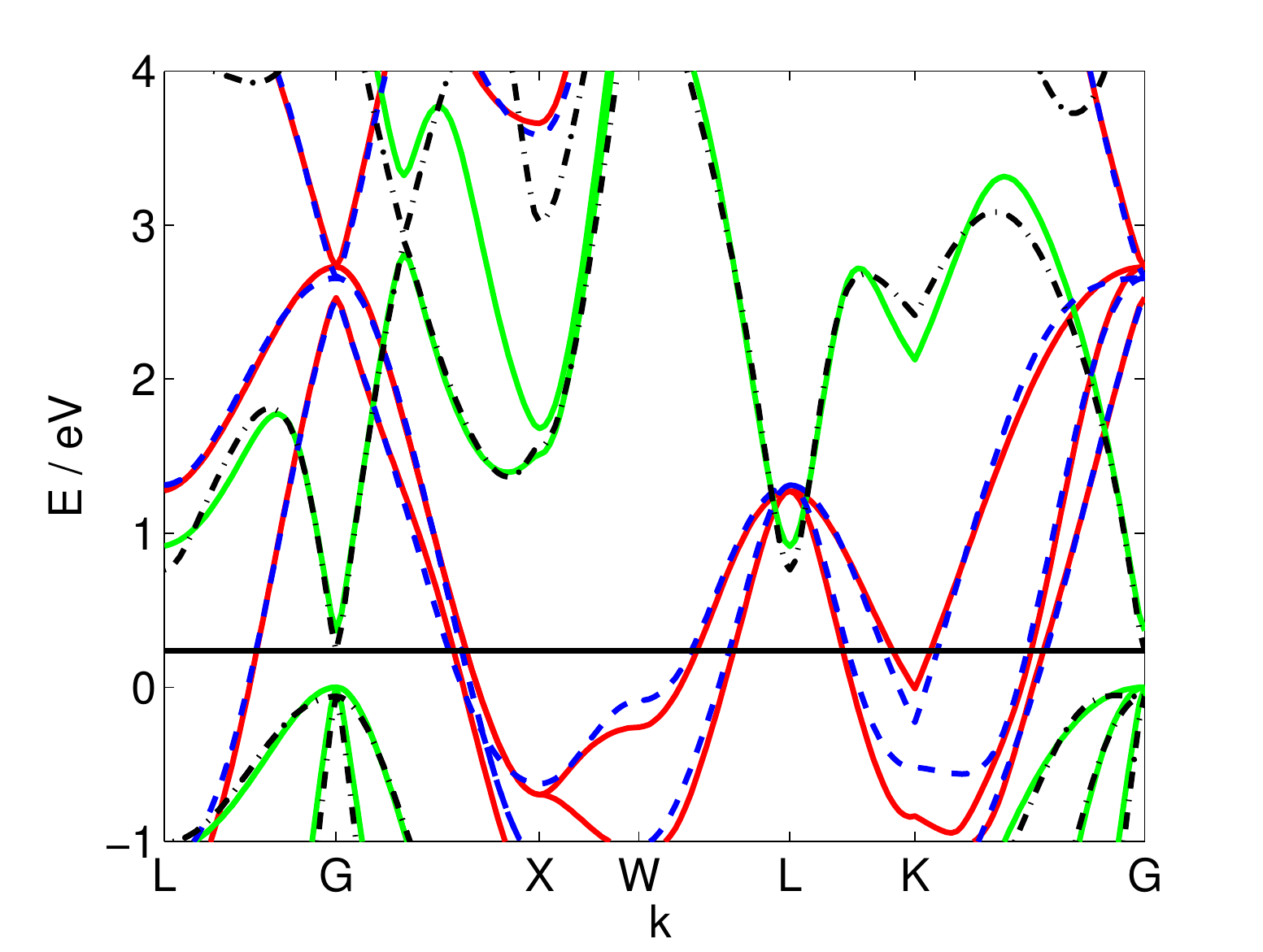}
 \caption{(Color online) Scenario (B): majority spin CrAs ab-initio band structure (red solid line), CrAs ETB-fit (blue dashed line), un-scissored GaAs band structure (green solid line) and GaAs ETB fit (black dashed line).
The Fermi energy $E_F = 0.01$ eV above the conduction band minimum of GaAs is indicated
by the horizontal solid line.} \label{fig:fitb}
\end{figure}

\begin{figure}[h!]
 \includegraphics[width = 80mm]{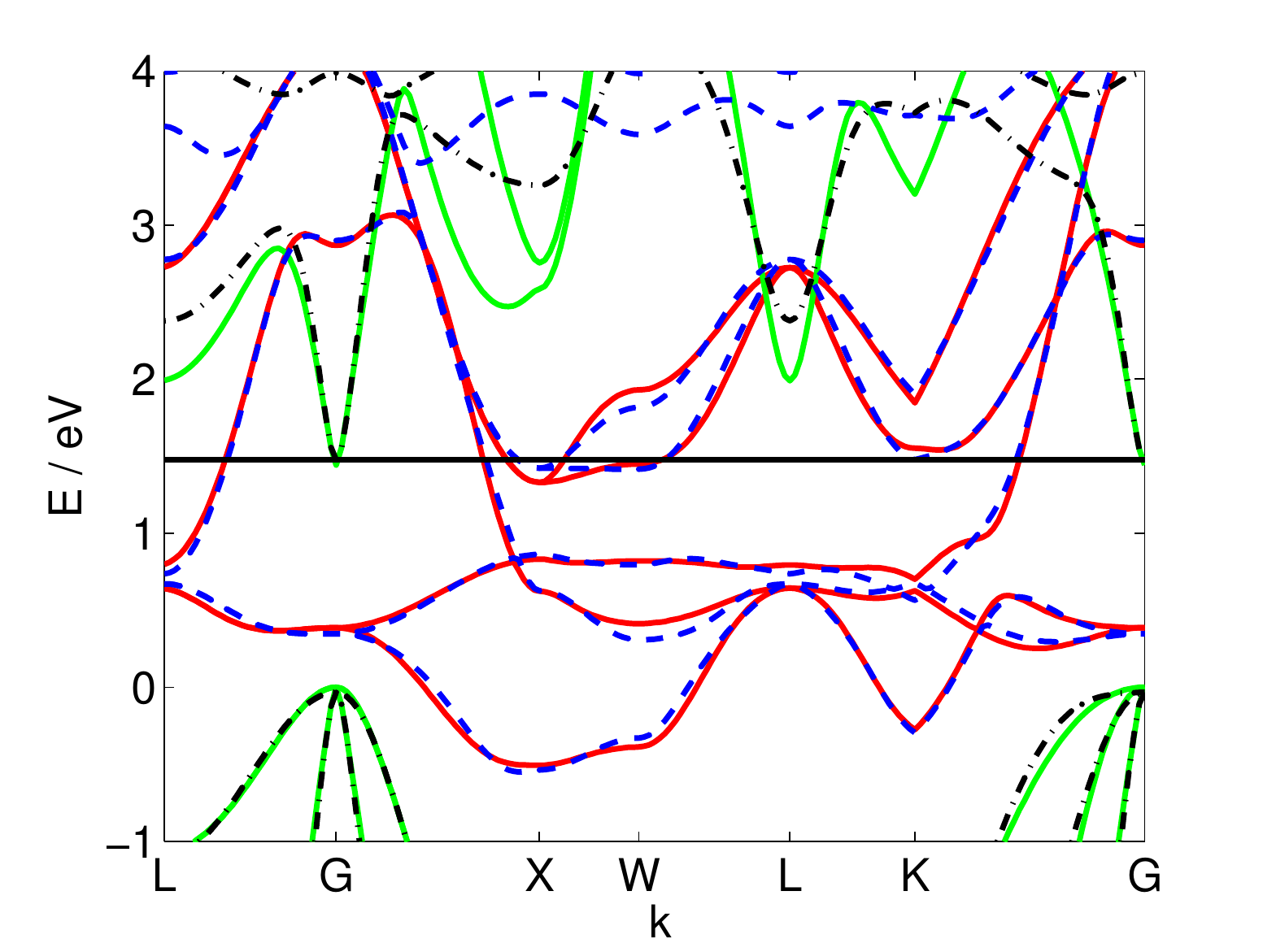}
 \caption{(Color online) Scenario (C): minority spin CrAs ab-initio band structure (red solid line), CrAs ETB-fit (blue dashed line), scissored GaAs band structure (green solid line) and GaAs ETB fit (black dashed line).
The Fermi energy $E_F = 0.01$ eV above the conduction band minimum of GaAs is indicated
by the horizontal solid line.} \label{fig:fitc}
\end{figure}

\begin{figure}[h!]
 \includegraphics[width = 80mm]{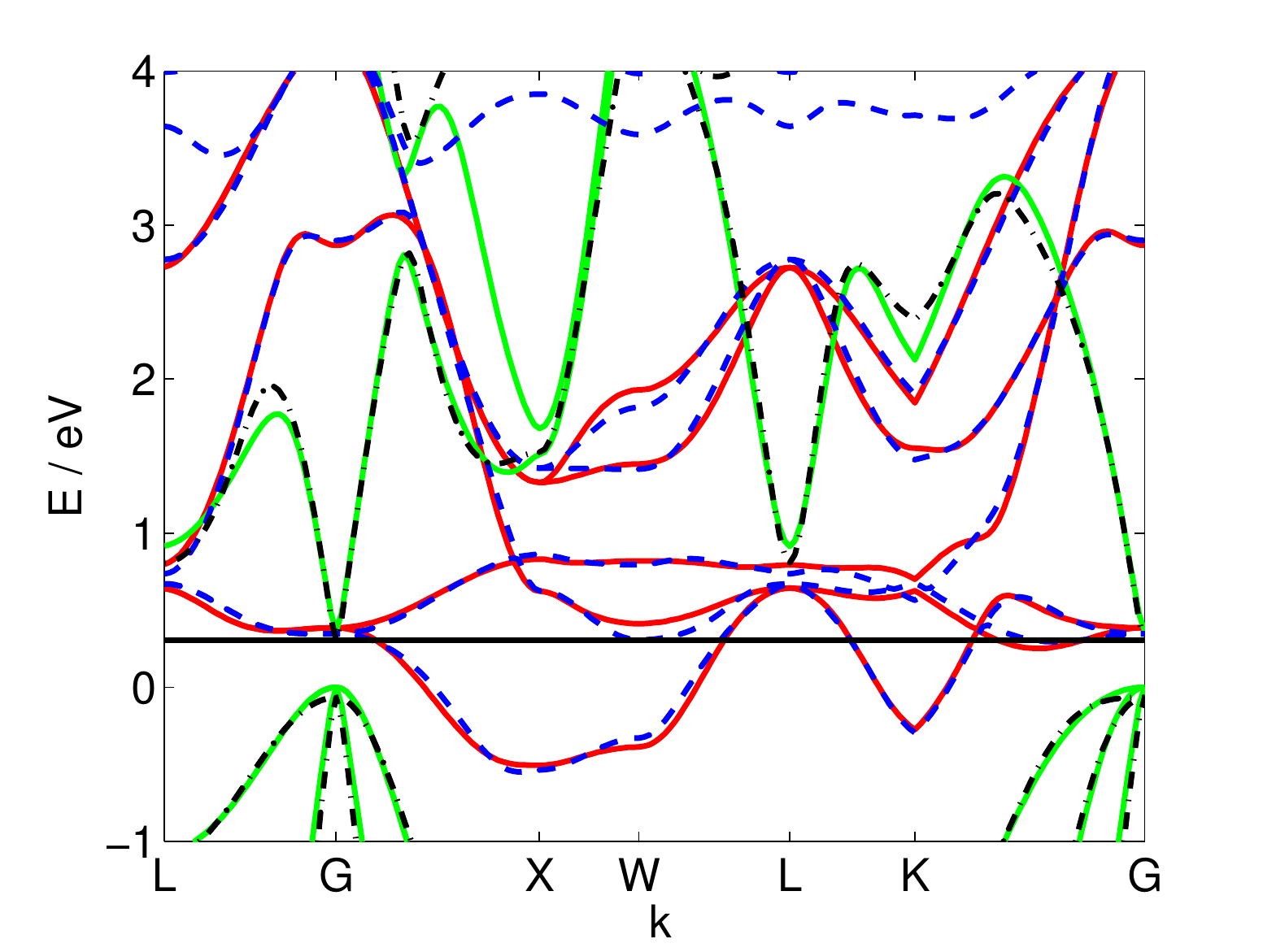}
 \caption{(Color online) Scenario (D): minority spin CrAs ab-initio band structure (red solid line), CrAs ETB-fit (blue dashed line), un-scissored GaAs band structure (green solid line) and GaAs ETB fit (black dashed line).
The Fermi energy $E_F = 0.01$ eV above the conduction band minimum of GaAs is indicated
by the horizontal solid line.} \label{fig:fitd}
\end{figure}

\section{Steady-state transport} \label{sec:trans}

The I-V-characteristics of the heterostructure is calculated within a non-equilibrium Green's function approach which has been adapted from our recent study of GaMnAs-based heterostructures, to which we refer for further details and references.\cite{ertler10,ertler11,dicarlo2003,stovneng94} The  Hamiltonian of the GaAs - CrAs - GaAs heterostructure is obtained by performing a partial Wannier transformation from the wave vector $k$ to $(x, k_\|)$, where $x$ denotes the [1,0,0] growth direction of the crystal and $k_\|$ is the in-plane (parallel) k-vector \cite{stovneng94},  
\be
\ket{nbjk_\|} = \frac{1}{L_{BZ}} \int \mathrm{d}k_x \exp \left ( - i k_x j \frac{a_\perp}{4} \right ) \ket{nbk},
\ee
where $L_{BZ} = \frac{8 \pi}{a_\perp}$ with  $j$ labeling the atomic layer. The resulting one-particle Hamiltonian is of the single-particle form
\bea
H(k_{\|}) & = & \sum_{i,\sigma} \varepsilon_{\sigma,\sigma}^{(i)} c_{i,\sigma}^\dagger(k_{\|}) c_{i,\sigma}(k_{\|}) \notag \\
 && +\sum_{i,\sigma\sigma'}t_{\sigma\sigma'}^{(i)}(k_{\|}) c_{i+1,\sigma}^\dagger(k_{\|}) c_{i,\sigma'}(k_{\|}) + \mathrm{h.c.},
\eea
with $c_{i,\sigma}^\dagger(k_{\|})$ [$c_{i,\sigma}(k_{\|})$] denoting the creation (annihilation) operator for site $i$ and orbital $\sigma$. $\varepsilon_{\sigma\sigma'}^{(i)}(k_{\|})$ and $t_{\sigma\sigma'}^{(i)} (k_{\|})$, respectively,  are onsite and hopping matrix elements. 
The semi-infinite GaAs ``leads" are taken into account by evaluating the associated self-energies and feeding them into  the system's Dyson equation.\cite{datta}  The surface Green's functions are obtained  with the help of an algorithm suggested by Sancho et al. \cite{sancho85}. 
For each carrier type (majority and minority), the transmission function $T(E,k_\|)$ for total energy $E$ and in-plane momentum $k_\|$ is calculated via
\be
T(E, k_\|) = \tr{\Gamma_R G^R \Gamma_L G^A}.
\ee
Here, $G^{R/A}$ are the system's retarded (R) and advanced (A) Green's functions, $\Gamma_{L/R}$ are the coupling functions to the left (L) and right (R) GaAs leads and $\tr{\cdot}$ is the trace operation. We then compute the steady-state current $j(V_a)$ assuming local thermal equilibrium among the electrons injected from a particular contact using the standard expression from stationary scattering theory\cite{datta}
\be
j(V_a) = \frac{2 e}{h} \sum_{k_\|} \int \mathrm{d}E T(E,k_\|) \left [f_L(E) - f_R(E) \right ],
\label{jva}
\ee
with $e$, $h$, and $f_{L/R}$ denoting, respectively, the elementary charge, Planck's constant, and the Fermi-Dirac distribution function for the left and right electric contact. 
The applied voltage $V_a$ enters Eq. \eqref{jva} in two places: the transmission function $T(E,k_\|)$ and the difference in the quasi-Fermi levels between left and right contact. 
In order to cut computational cost,  we assume a linear voltage drop from the left to the right lead across the simulated structure. This implies a somewhat artificial relationship between the electric field across the structure and the applied bias.\cite{poetz92,poetz89} 
In principle, both an effective single-particle potential, an effective exchange splitting, and a self-consistent treatment of charge injection can be implemented into the present model.\cite{ertler11}
However, while providing a significant reduction in computation time, omission of self-consistency may not significantly reduce the quality of our results. Note that the Fermi-Dirac distributions of the GaAs contacts provide the sole temperature dependence in the current model since a temperature dependence of the electronic structure is not considered here. For a discussion of the latter, we refer to recent work.\cite{chioncel11}

In what follows we present results for the I-V characteristics of GaAs/(CrAs)$_\ell$/GaAs heterostructures for $\ell= 4, 6,8,10$.  While thin layers of fcc CrAs may be easier to realize experimentally, thicker layers thereof are 
described more realistically within our approach.  
The free carrier density in the n-doped  GaAs regions is about $4.5 \times 10^{17}$ cm$^{-3}$ at $T = 300$ K ($7.9 \times 10^{16}$ cm$^{-3}$ at $T = 0$ K), with the quasi-Fermi level held constant at $10$ meV above the conduction band edge. The  applied bias was varied between zero and $0.2$ V.  Results for scissored and un-scissored GaAs, respectively,  and $ \ell=10$ are shown in Figs. \ref{fig:iv1} and \ref{fig:iv2} (mind the semi-logarithmic plot).     The overall features of the I-V characteristics agree for both cases: the majority current density clearly dominates over the minority current density and this, in most bias regions, by several orders of magnitude.  However, spin-filtering is more pronounced for the scissored GaAs model.  While the majority current is rather insensitive to scissoring, the minority current density is not (we believe that the small oscillations for the $77$ K minority case near $1.5$ V in Fig.
 \ref{fig:iv1} are of numerical origin).  The reason is found by inspection of Figs. \ref{fig:esmaj} and \ref{fig:esmin}.  
It shows that, for un-scissored GaAs and low applied bias, there is  a resonance between the GaAs conduction band minimum (dashed line) 
and CrAs-associated bands near the $\Gamma$-point.  Near the $\Gamma$-point these bands are rather flat and so the group velocity  is almost zero.   Under moderate bias, however,  these bands are moved further into resonance (to regions with higher group velocity) with the conduction band of GaAs at the emitter side, and the minority current rises steeply with applied bias.   

\begin{figure}[h!]
 \includegraphics[width = 80mm]{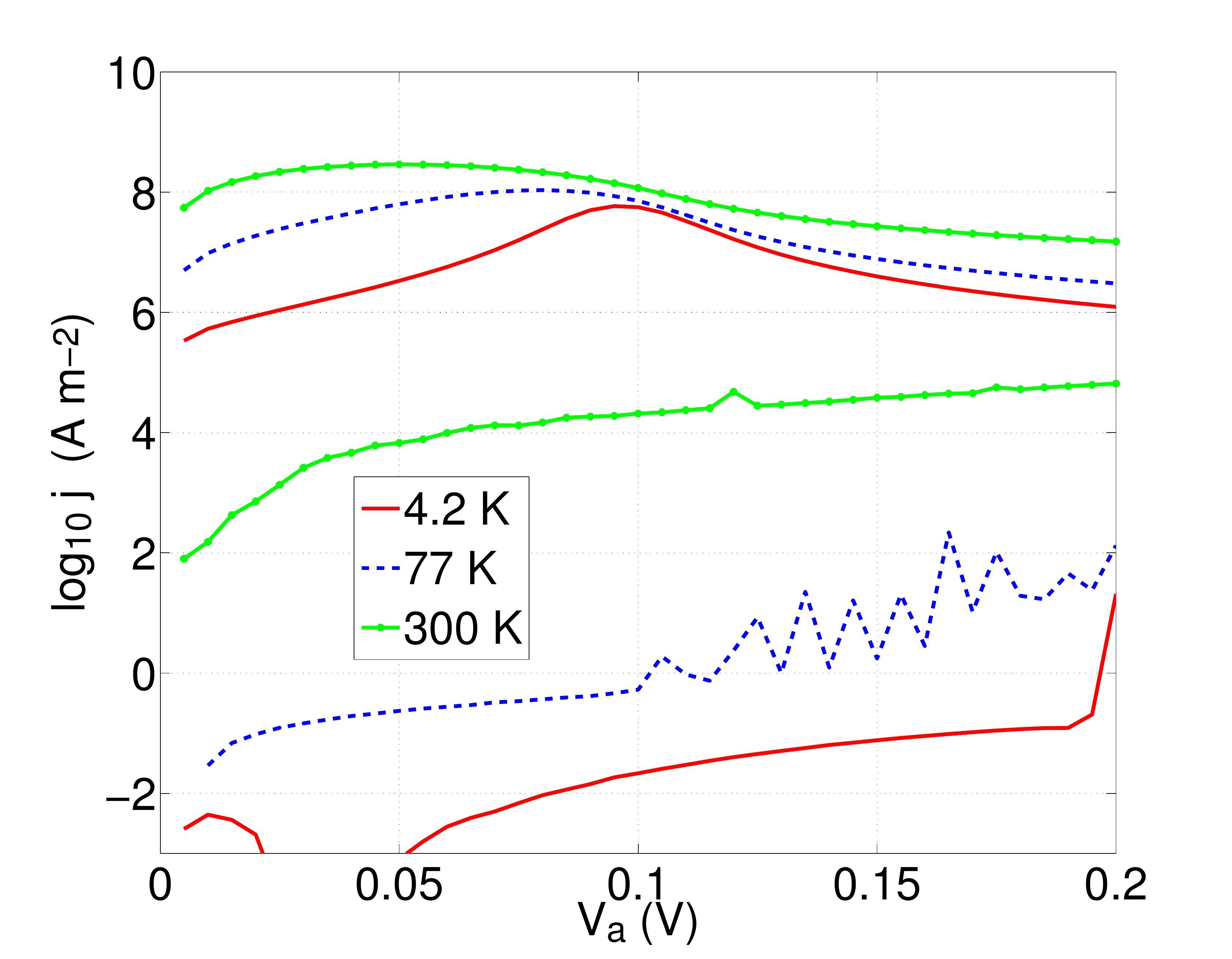}
 \caption{(Color online) Spin-resolved current-voltage characteristics for GaAs/(CrAs)$_{10}$/GaAs and case (A) and (C) (scissored GaAs) for different temperatures
($T = 4.2$~K, $T = 77$~K, and $T = 300$~K). The majority spin current clearly dominates the minority spin current for all voltages and temperatures. } \label{fig:iv1}
\end{figure}

\begin{figure}[h!]
 \includegraphics[width = 80mm]{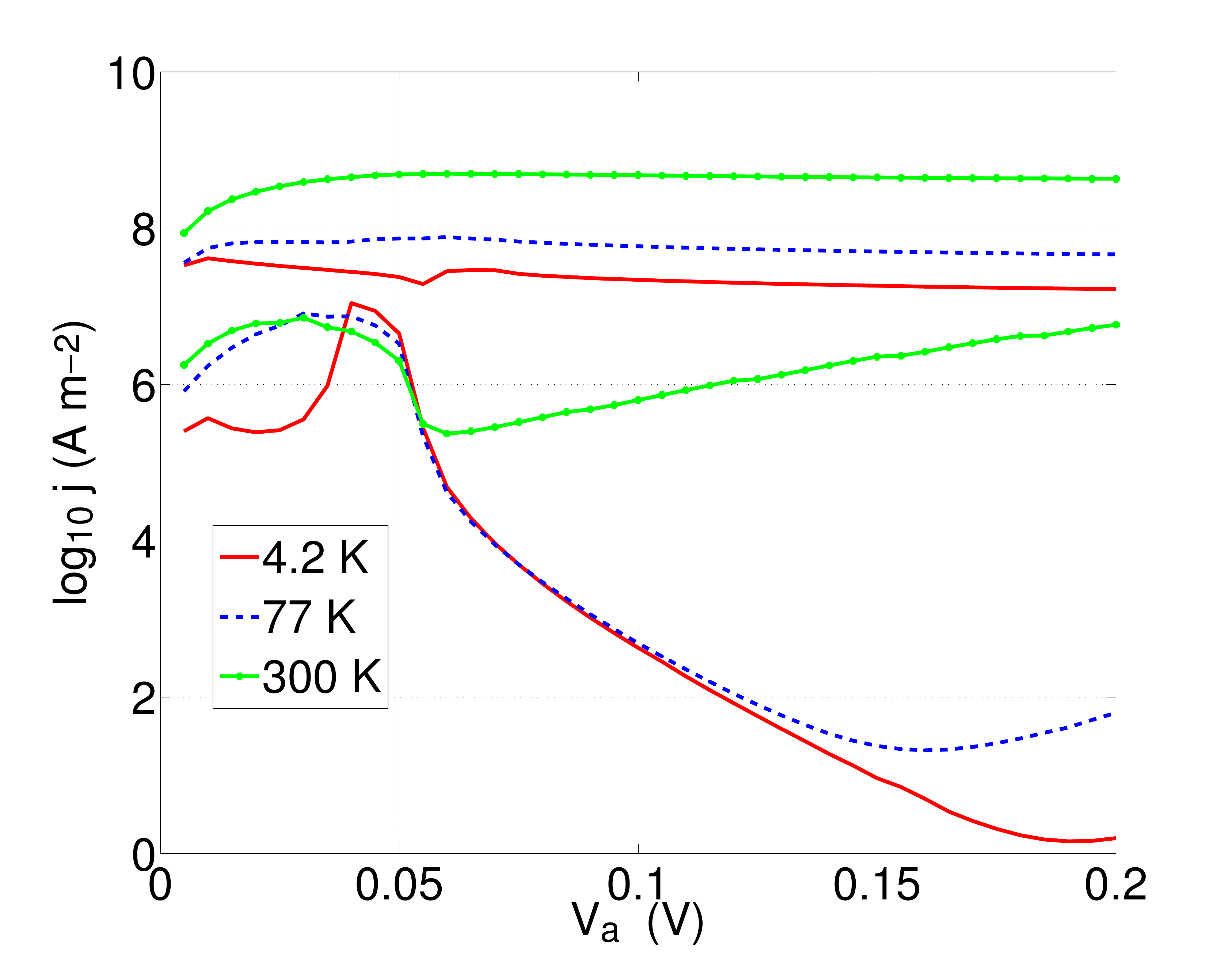}
 \caption{(Color online) Spin-resolved current-voltage characteristics for GaAs/(CrAs)$_{10}$/GaAs and case (B) and (D) (un-scissored GaAs) for different temperatures
($T = 4.2$~K, $T = 77$~K, and $T = 300$~K). The majority spin current clearly dominates the minority spin current for all voltages and temperatures.} \label{fig:iv2}
\end{figure}

The current spin polarization $P(V_a)$ as a function of applied voltage $V_a$ is defined as
\be
P(V_a) = \left \vert \frac{j_{\rm maj}(V_a) - j_{\rm min} (V_a)}{j_{\rm maj}(V_a) + j_{\rm min} (V_a)} \right \vert.
\ee
Here, $j_{{\rm maj/min}}$ refers to the majority and minority spin current density, respectively. In Figs. \ref{fig:spinpol1} and \ref{fig:spinpol2}, respectively, we display the computed current spin polarization for scissored and un-scissored GaAs  and the three different temperatures discussed above.

\begin{figure}[h!]
 \centering
 \includegraphics[width = 80mm]{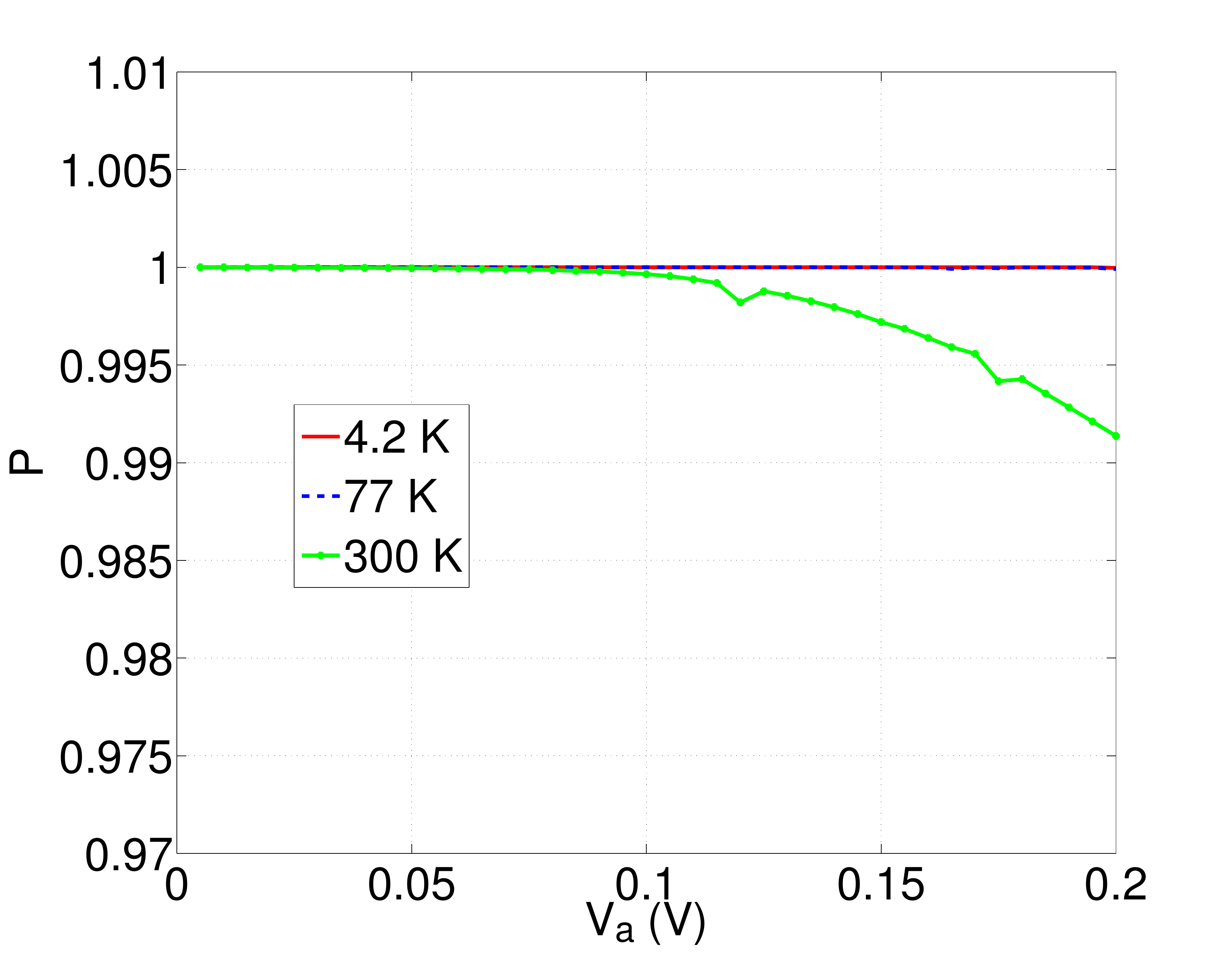}
 \caption{(Color online) Current spin polarization $P(V_a)$ for scissored GaAs for different temperatures
($T = 4.2$~K, $T = 77$~K, and $T = 300$~K).} \label{fig:spinpol1}
\end{figure}

\begin{figure}[h!]
 \centering
 \includegraphics[width = 80mm]{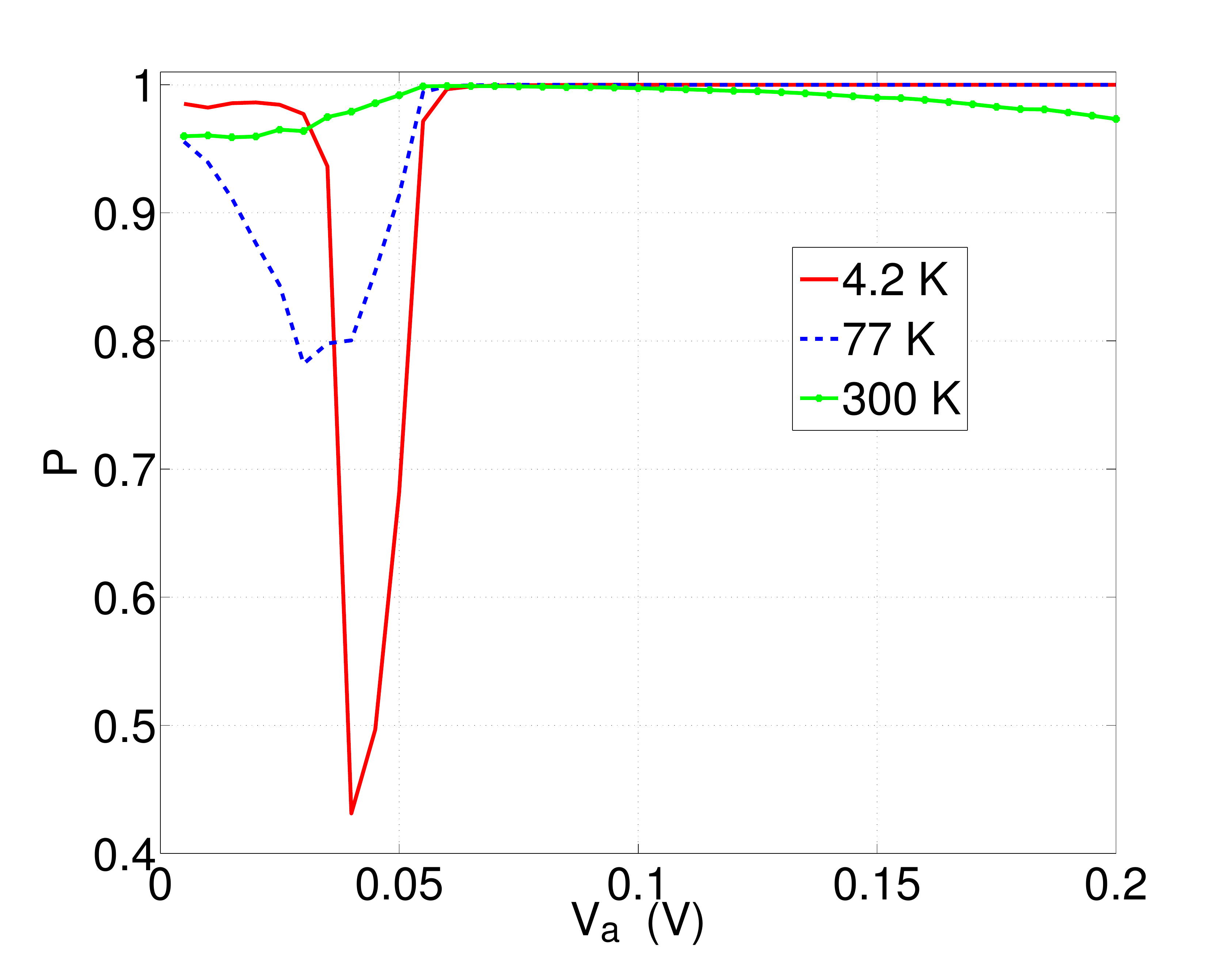}
 \caption{(Color online) Current spin polarization $P(V_a)$ for un-scissored GaAs for different temperatures
($T = 4.2$~K, $T = 77$~K, and $T = 300$~K).} \label{fig:spinpol2}
\end{figure}

In Fig. \ref{fig:layers} we show the I-V characteristics as obtained for different layers thicknesses, i.e. $($GaAs$)_m$/$($CrAs$)_{\ell}$/$($GaAs$)_m$, where $\ell = 4,6,8,10$ and $m = 5$. The results shown in Fig. \ref{fig:layers} stem from simulations in which the number of layers of GaAs to the left and the right of CrAs was set to $m = 5$ and kept constant, i.e. the electric field across the CrAs layer at a given voltage increases with decreasing layer thickness.  This trend follows the actual physical trend within the device and that of a self-consistent model.  Moreover, we note that the actual form of the computed I-V characteristic is only slightly changed if our simulations are performed under equal-electric-field conditions, i.e. $m = 10 - \frac{\ell}{2}$ because the form of the bands involved in the transport of majority carriers (determining their transmission coefficient) is very robust under a slight change of the energy offset, see Figs. \ref{fig:fita} and \ref{fig:fitb}.

From Fig. \ref{fig:layers} we clearly observe non-Ohmic behavior which is due to the rather complicated electronic structure involved in the transmission probability. We observe, for instance, that the absolute value of the current transmitted through a structure consisting of six layers CrAs is higher for all voltages than when  transmitted through four layers. These I-V characteristics indicate that spin-filtering should also be realizable with very thin structures of CrAs, see Fig. \ref{fig:layerspol}, which might be easier to fabricate.  For n=4, the minority current shows nonlinearities which we attribute to resonant transport mediated by states which, in the bulk, give rise to the bands discussed above. Nevertheless, it has to be kept in mind that the systematic error of our approach is larger for very thin structures because (i) the modeling of the CrAs  layers is based on the Hamiltonian of bulk ZB CrAs and (ii)  effects from the interface will become more important for thin layers. Nevertheless, 
spin-filtering should be observable.

\begin{figure}
 \centering
 \includegraphics[width = 80mm]{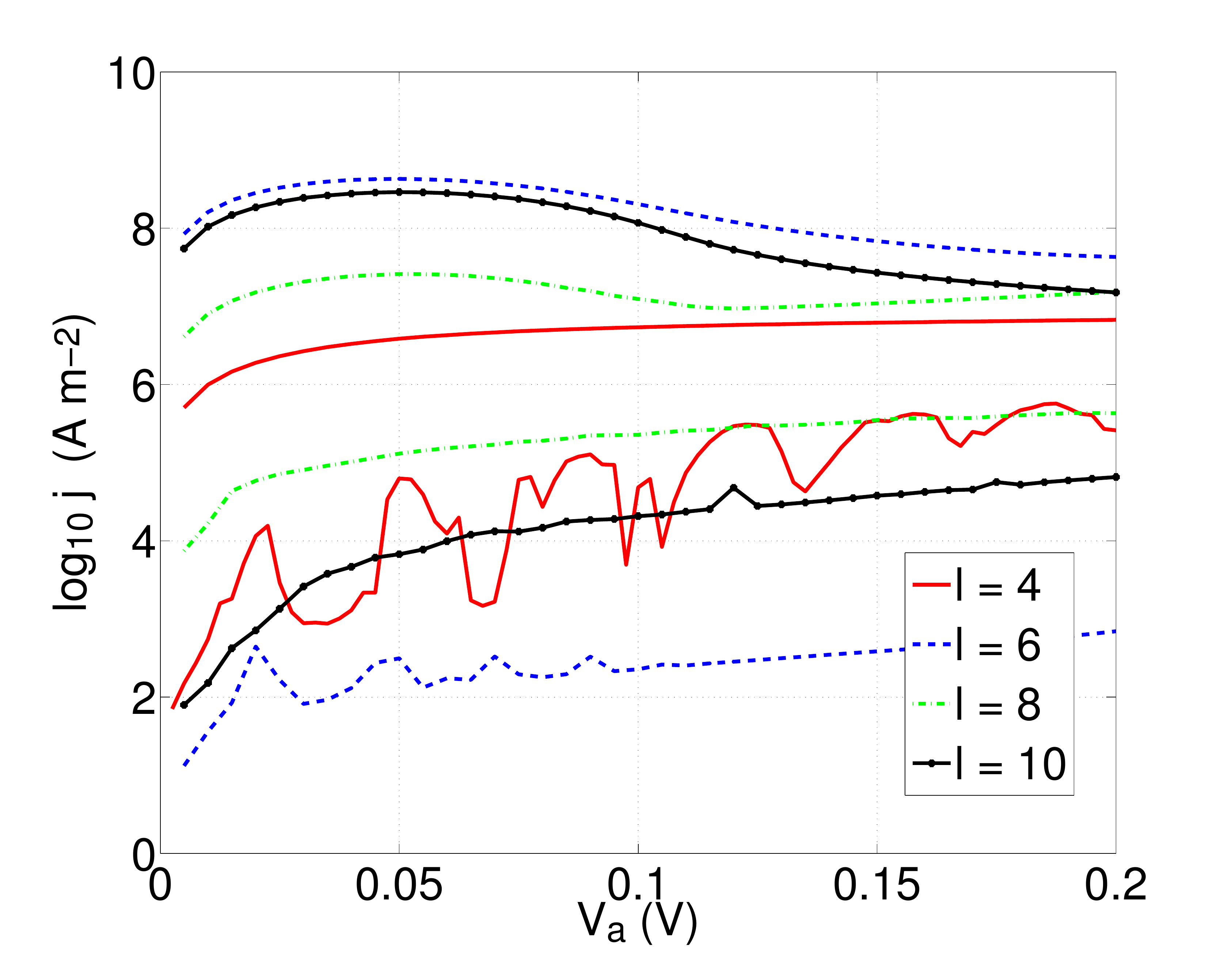}
 \caption{(Color online) I-V characteristics for different layer thicknesses $\ell = 4,6,8,10$.} \label{fig:layers}
\end{figure}

\begin{figure}
 \centering
 \includegraphics[width = 80mm]{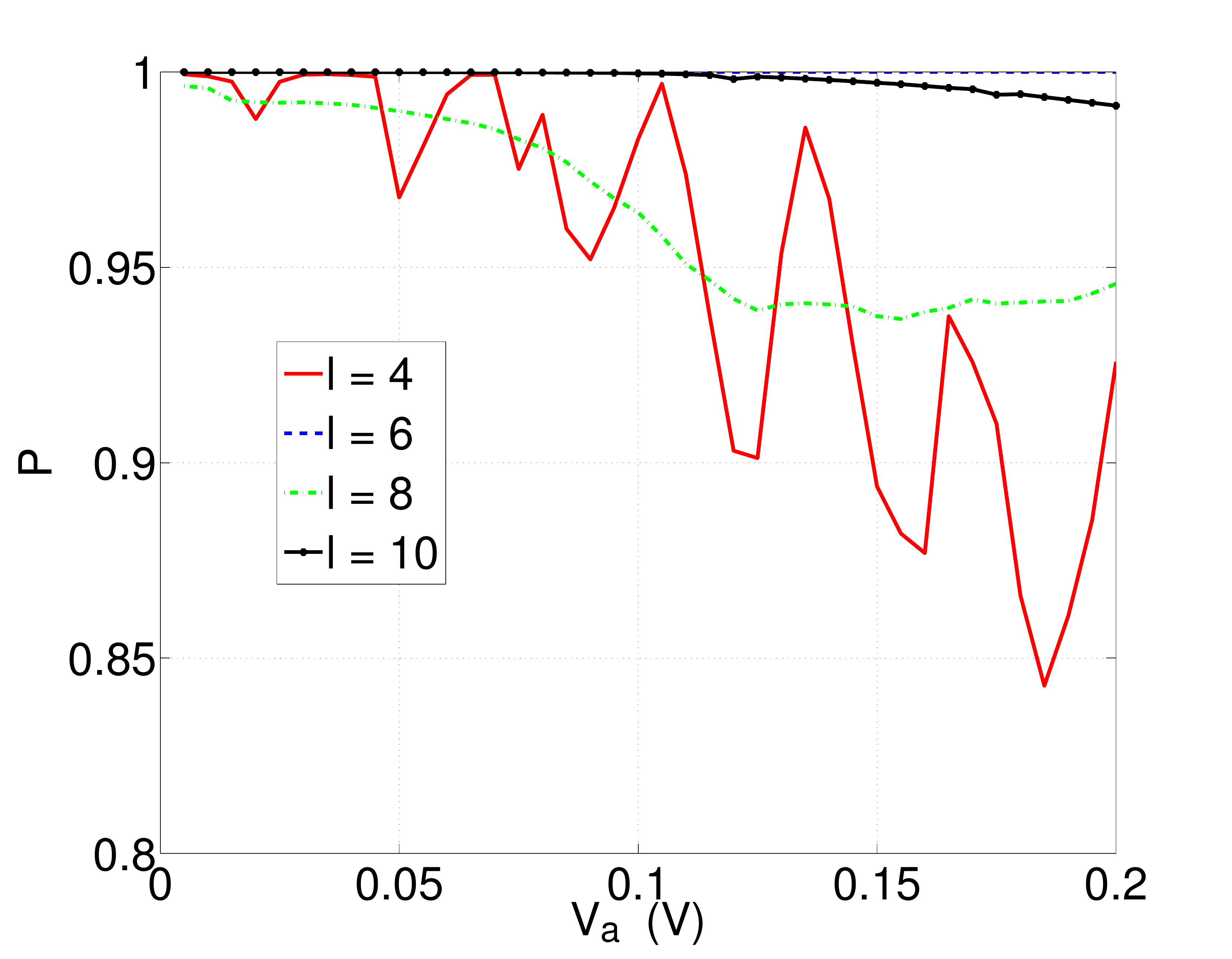}
 \caption{(Color online) Current spin polarization for different layer thicknesses $\ell = 4,6,8,10$.} \label{fig:layerspol}
\end{figure}

% Since the 

\section{Summary, discussion, conclusions, and outlook} \label{sec:conc}

We have performed a model study of transport in CrAs / GaAs heterostructures which is based on the assumption that sufficiently thin layers of CrAs can be grown in between a GaAs substrate in lattice matched fashion. The bulk electronic structure of fcc GaAs, fcc and tetragonal CrAs, as well as lattice matched single [1,0,0] GaAs / CrAs heterointerfaces were calculated within an LSDA LMTO model and used to determine the band offsets between the two materials for minority and majority carriers.  As a remarkable result, we find a (local) total energy minimum for a hexagonal bulk CrAs unit cell, when the transverse lattice constant is held fixed at the bulk GaAs value  $a_{\rm GaAs} = 5.65 \AA$ and the longitudinal [1,0,0]  lattice $a_\perp$ constant is varied.  The minimum is found for $a_\perp \approx 0.98 a_{\rm GaAs}$.  Although this local equilibrium unit cell of CrAs is tetragonal rather than cubic, the systematic errors introduced by assuming a perfectly lattice matched fcc crystal structure at $a_{\rm 
GaAs}$ is found to be negligible. For a lattice constant of $a_{\rm CrAs} = 5.65 \AA $ we find that fcc CrAs is a half-metal, with zero gap for one spin orientation (majority carriers) and a gap of $1.8$ eV at the $X$ point for the other (minority carriers). The computed band offset for a lattice matched [1,0,0] hetero-interface between the two materials is found to be about $ 0.5-0.6\pm 0.2$ eV.  The latter implies an alignment of the gap region of minority CrAs with the {\it central} region of the uppermost valence bands of GaAs.  For spin-filtering, therefore, it is not important whether the sheet of CrAs  is half-metallic or not.  

The ab-initio spin-dependent electronic band structures are mapped onto  a ETB model which is used to construct the effective Hamiltonian of the n-GaAs/CrAs/n-GaAs  heterostructures consisting, respectively,  of $\ell= 4,6,8,10$ mono-layers of CrAs.   This down-folding was constrained by the requirement that As ETB onsite parameters for a given spin orientation be constant throughout the system, thereby, eliminating the need for the introduction of ad-hoc ETB parameters at the heterointerface. 
The current response for majority and minority carriers is obtained within a non-equilibrium Green's function approach. We consider carrier injection from n-doped GaAs and our calculations show efficient spin-filtering over a wide parameter range, in particular, regarding the precise band alignment between the GaAs conduction band edge with the CrAs bands, temperature, and layer thickness.   Spin-polarization of up to $99$ percent, as well as room-temperature spin-filtering, is predicted within this model.  

% While a number of potential improvements to the present theoretical approach, such as a more realistic account of correlation, inclusion of the spin-orbit interaction, a self-consistent treatment of transport, etc.,  
% can readily be listed, the main problem currently lies in the fabrication of  fcc hetrerostructures containing  layers of  transition metal compounds, such as MnAs or CrAs, and  conventional semiconductors, such GaAs.  While for MnAs there seems to be little hope left, apart for Mn delta-doped GaAs structures and strained fcc MnAs quantum dots on GaAs, experimental evidence for fcc CrAs layers on GaAs substrates still seems to be controversial.  It is hoped that these promising theoretical results regarding high spin-polarization encourage the materials growth and experimental physics community in the study of semiconductor heterostructures 
% containing transition metal compounds such as CrAs, MnAs, or VAs.

A number of potential improvements to the present theoretical approach, such as a more realistic account of correlations, the inclusion of the spin-orbit interaction, a self-consistent treatment of transport, a more detailed inclusion of the interface, scattering, etc., can readily be listed and be addressed in future studies.   However, we are convinced that the main statement of this publication, i.e. that lattice matched CrAs/GaAs heterostructures are strong candidates for room-temperature spin-filters,  is not affected by these details.  A potential source for major corrections could be a strong bias dependence of the exchange interaction, i.e. a bias anomaly, in CrAs, similar to the one predicted and reported for heterostructures containing GaMnAs.\cite{ohya11,ertler12}

At this point, however, we believe that experimental assistance is essential to make further progress.  It is our hope that with the present results  we can stimulate a renewed interest in the fabrication of  fcc hetrerostructures containing  layers of  transition metal compounds, such as CrAs, MnAs, or VAs, and  conventional fcc semiconductors.  For MnAs, apart from Mn $\delta$-doped GaAs structures and strained fcc MnAs quantum dots on GaAs, the growth in the  fcc or tetragonal phase apparently has not been successful.   Evidence for fcc (or tetragonal) CrAs layers on GaAs substrates still seems to be controversial. It is hoped that these promising theoretical results regarding the CrAs electronic structure, the favorable band alignment with GaAs, and the predicted high spin-polarization in charge transport encourage the materials growth and experimental physics community in a continued study of semiconductor heterostructures containing transition metal compounds.

\section{Acknowledgments}

The authors thank R. Hammer, M. Aichhorn, and E. Arrigoni  for fruitful discussions. This work was supported financially  by FWF project P221290-N16.

\appendix

\begin{widetext}
\section{tight-binding parameters}

\begin{center}
\begin{table}[h!]
\centering
 \caption{tight-binding parameters for bulk CrAs majority and minority spin and the respective GaAs parameters. The anion onsite energies are indicated by the number $1$ while the cation is labeled by the number $2$. For further notations see \cite{jancu98}.} \label{tab:fitpara}
\centering
\begin{tabular}{c || c | c | c || c | c | c}
& CrAs maj & GaAs maj (A) & GaAs maj (B) & CrAs min & GaAs min (C) & GaAs min (D) \\
\hline
\hline
 $E_{s1}$ &     2.6574 & 2.6574&2.6574&4.5703&4.5703&4.5703\\
 $E_{s2}$ &     4.3837 &4.1719&-11.8119&1.0108&5.5583&-16.4426\\
 $E_{p1}$ &     1.2291 &1.2291&1.2291&2.0184&2.0184&2.0184\\
 $E_{p2}$ &    13.8494 &26.2033&11.8069&-6.5536&5.4491&27.2734\\
 $E_{d11}$ &     8.2985 &8.2985&8.2985&9.9412&9.9412&9.9412\\
 $E_{d12}$ &     7.4409 &7.4409&7.4409&6.3685&6.3685&6.3685\\
 $E_{d21}$ &    -1.3823 &15.6055&14.1494&4.5335&4.5941&9.9138\\
 $E_{d22}$ &    -1.9278 &14.1638&10.6150&0.8616&5.5512&14.2845\\
 $E_{s^*1}$ &    18.5203 &18.5203&18.5203&-0.6061&-0.6061&-0.6061\\
 $E_{s^*2}$ &     1.8607 &10.4258&9.3580&9.0261&-2.4055&16.4261\\
\hline
 $(ss\sigma)$ &    -1.1307 &15.6358&8.1308&1.3311&0.2533&-0.2897\\
 $(s1p2\sigma)$ &     0.9972 &-2.2562&0.5449&2.9666&0.2245&-1.8973\\
 $(s2p1\sigma)$ &    -0.1885 &7.2869&5.2848&-1.4504&0.7586&7.7539\\
 $(pp \sigma)$ &     3.2695 & -4.9820&3.2030&-3.5148&-2.3875&-0.9973\\
 $(pp \pi)$ &    -1.9484 &2.7602&-2.0813&-0.1283&1.3346&2.9221\\
 $(s1d2\sigma)$ &    -3.3586 &-4.3212&3.9507&0.4038&-0.6059&-1.3790\\
 $(s2d1 \sigma)$ &     3.1659 &-3.3578&-1.4149&2.4298&0.6413&-4.6244\\
 $(p1d2 \sigma)$ &   -1.1220 &-0.2759&0.1416&0.5525&2.7231&3.8984\\
 $(p1d2 \pi)$ &     0.8269 & 2.5960&2.5355&0.5372&0.3951&2.0218\\
 $(p2d1 \sigma)$ &    0.1535 &2.1035&0.4901&3.7059&-0.2031&0.7540\\
 $(p2d1 \pi)$ &   -1.9975 &-3.6283&-2.8348&3.0276&0.9990&1.6321\\
 $(dd \sigma)$ &   -2.1444 &-1.4358&0.7386&-2.1594&0.1633&-0.3069\\
 $(dd \pi)$ &   -0.5411 &0.1829&-0.7089&0.3193&-0.1910&-1.7396\\
 $(dd \delta)$ &    0.3357 &-0.2430&-1.2726&0.6818&0.4668&2.9257\\
 $(s1s^*2\sigma)$ &   -0.9709 &2.3385&7.1707&0.0642&-0.0772&1.9910\\
 $(s2s^*1 \sigma)$ &    3.3994 &-5.0239&0.4364&3.0872&-0.2349&12.4053\\
 $(s^*1p2 \sigma)$ &    -2.7234 &-4.8008&1.7586&-2.5451&0.5351&16.9906\\
 $(s^*2p1 \sigma)$ &     4.5299 &5.7515&0.6181&0.6516&4.6082&4.5138\\
 $(s^*1d2 \sigma)$ &     2.8069 &-1.7050&3.4076&1.0414&-3.8518&12.5601\\
 $(s^*2d1 \sigma)$ &     0.2943 &-2.6707&-1.6950&-2.1579&2.2004&-3.5301\\
 $(s^*s^* \sigma)$ &     3.8433 &-3.8555&0.3662&1.4586&0.7470&-5.2658\\
 \end{tabular}
\end{table}
\end{center}
\end{widetext}

\bibliographystyle{unsrt}
\bibliography{report}

\end{document}